
\baselineskip=18pt plus 2pt minus 1pt
\magnification=1200
\hsize=5.7truein
\vsize=8.4truein
\voffset=24pt
\hoffset=.1in
\centerline{\bf
Non-Fermi-Liquid-Like Behavior in The Two-Dimensional $t-J$ Model }
\vskip .3in
\centerline{Y. C. Chen$^1$ and T. K. Lee$^2$ }
\centerline{$^1$Dept. of Physics, National Tsing Hua Univ., Hsinchu, Taiwan}
\vskip .15in
\centerline{$^2$Center for Stochastic Processes in Science and Engineering}
\centerline{and}
\centerline{Department of Physics}
\centerline{Virginia Polytechnic Institute and State University}
\centerline{Blacksburg, VA 24061 USA}
\vskip .15in
\vskip .3in
The spin, charge and pairing correlation functions for the ground state of
two-dimensional $t-J$ model are calculated by using
the power method which projects out the ground state from a variational
wave function.
``Exact''  results are obtained for $10\times10$
and $8\times8$ lattices with
low particle densities.
 The results are surprisingly similar to that of one-dimensional $t-J$
model. The phenomenology is almost perfectly correlated with the concept of
Tomonaga-Luttinger liquid and  it is incompatible with traditional Fermi liquid
theory.


\vfil\eject

It has been well accepted now that the normal state of the high temperature
superconductors is not an ordinary Fermi liquid (FL) [1].
For the last several years Anderson [2,3] has been promoting the idea of
a Tomonaga-Luttinger liquid (TLL) [4].
TLL is shown by exact theories [5] and numerical studies
[6,7] to be the ground state of the Hubbard model and $t-J$ model
 in one dimension. But until now there is no evidence
that the $t-J$ model or any other model in two dimensions will indeed have such
a ground state.

In this letter we present numerical results for the ground-state
correlation functions calculated by the power method [7] in the low
particle-density region. The
calculations are done on an $8\times8$ square lattice and on a
$10\times10$ lattice. The overall data have a surprisingly similarity with
what we have found in one-dimension [7].
The variations of the
spin, charge and pairing correlation functions with the coupling
constant $J/t$ can be easily understood by borrowing the concept of
the correlation exponent $K_{\varrho}$ in one dimension. TLL provides
a consistent interpretation of all the observed features
 in these
correlation functions [3].

Exact results in the ground
state of the $t-J$ Hamiltoian have been obtained
by diagonalizing small lattices such as 16 or 18 sites [8,9].
Various variational approaches have been used to study
larger lattices [10-11]. Recently, Valenti and Gros [12] generalized
a trial wave function used by Hellberg and Mele [13] in one dimension
 to study the two-dimensional $t-J$ model. This wave function
 (HMVG) is shown to be a TLL in two dimensions.
Due to the nature of the variational approach it is difficult to assess
how close the variational state is to the true ground state.

Here we adopt a ground-state projection method
that systematically improves the
variational wave function and
provides quantitative information about the ground state.
 This method which we shall
refer to as the power method is a simplified version of the Green's function
Monte Carlo method [14,15]. The mathematical idea is quite simple.
Given a wave function $\left|\psi\right>$ that is not orthogonal to
the ground state of a Hamiltonian $H$,  applying the
operator $(W-H)^p$  to $\left|\psi\right>$ will project out the ground
state as the power $p$ approaches infinity. The constant $W$ is chosen
such that all the excited state with energy $E_i$ satisfies the relation
$|W-E_i|<|W-E_g|$, where $E_g$ is the ground state energy. For the $t-J$
Hamiltonian, it is sufficient to choose $W=0$.
The numerical technique we use to implement the power method is
a combination of the variational Monte Carlo (VMC) method
and the Neumann-Ulam matrix method [16]. Specific details of our method
can be found in reference 7.

 All the data presented in this paper is
obtained by averaging ten to twenty independent groups. Each group usually
consists of one to two thousands starting configurations determined by
the chosen trial wave
function.  Each starting configuration then would produce a couple
hundred paths. The calculations are all done in workstations with
average running time about a day or less. Obviously,
the amount of computer time required
is very sensitive to the size of the
system, the choice of the initial trial wave function and how many
powers required. In two dimensions this method usually cannot
be carried out for very large powers because of the Fermion ``sign''
problem [17]. Fortunately, for low particle density the sign
problem is not severe. For power equals to twenty, the negative terms are
only about six percent of the total.

This method has been
successfully applied to study the $t-J$ model
in one dimension [7,18].
The calculated results
of the static structure factors of
longitudinal spin, $S(k)$; of charge, $D(k)$; and of pairing, $P(k)$,
have a power-law
decay in real-space as predicted by the TLL [5,6].
All correlation functions are determined by a
dimensionless correlation exponent $K_{\varrho}$:
$$
 \left< S_z(r) S_z(0) \right>\sim A_0r^{-2}+A_1 cos(2k_Fr)
r^{-(1+K_{\varrho})},\eqno(1)
$$
$$
 \left< n(r) n(0) \right>\sim B_0r^{-2}+B_1 cos(2k_Fr)
r^{-(1+K_{\varrho})}+ B_2 cos(4k_Fr)r^{-4K_{\varrho}},\eqno(2)
$$
$$
 \left< \Delta^{\dag}(r) \Delta(0)\right>\sim
C_0r^{-(1+{1\over{K_{\varrho}}})}+C_1 cos(2k_Fr)
r^{-(K_{\varrho}+{1\over{K_{\varrho}}})},\eqno(3)
$$
where $r\gg1$ and
the singlet pairing operator $\Delta(r)=C_{r\uparrow}C_{r+1\downarrow}-
C_{r\downarrow}C_{r+1\uparrow}$. For $J/t<2$,
 $K_{\varrho}$ is less than 1. This is the repulsive TLL where
spin-density-wave (SDW) correlation dominates as reflected by a cusp in
$S(k)$ at $k=2k_F$. In this region pairing correlation is
suppressed and charge structure factor has a maximum at zone boundary
$k=\pi$.
For $J=2$,
 $K_{\varrho}$ is very close to 1 and we have the Fermi liquid behavior
for all the correlation functions [19]. When $J$ is increased above the
value of 2, pairing correlation becomes dominant with
 $K_{\varrho}>1$. This is the attractive TLL. P(k) has an
upward cusp at $k=0$ and a dip at $k=2k_F$. The larger the $J$ value the
stronger are these effects.

In order to make sure the final result is
independent of the choice of starting trial wave functions
we usually use
more than one wave function.
In two dimensions we use the Gutzwiller wave functions, GWF [20], and HMVG
[12].
GWF is just an ideal Fermi gas state satisfying the constraint of no double
occupancy at any site.
 The function HMVG is essentially of the
same form as GWF with a Slater determinant for up-spin electrons and
one for down-spin electrons. In addition to these two determinants
it has a long range correlation part between all the particles:
$\Pi_{i<j} \left|{\bf r}_i-{\bf r}_j\right|^{\nu}$, for nearest-neighbor
particles we chose $\nu=0$. Clearly when $\nu$ is positive HMVG gives
more weight for particles separated far apart, hence it represents
an effective repulsive interaction. On the other hand when $\nu$ is
negative an effective attractive interaction has been put into the
wave function. In one dimension a similar wave function
is shown by Hellberg and Mele [7,13] to nicely describe the repulsive and
attractive TLL.

 In figure 1(a) energy as a function of
power is plotted for 18 particles in a $10\times10$ lattice for $J=0.1$.
Energy is in unit of the hopping matrix element $t$.
The empty
circles are the results started with the GWF, while the filled circles
started with HMVG-$\nu=0.03$. Lines are guide for the eyes. The inset
shows the energy of 10 particles in 64 sites as a function of power p.
The empty circles are
results using GWF and filled circles for HMVG-$\nu=0.05$. Clearly
the same ground state energy is obtained by applying powers of $H$ to either
GWF or HMVG. The empty
squares in the inset of Fig. 1(a)
are results for $J=2$ started with GWF. GWF is again very
close to the true ground state for $J=2$ just as in one dimension
[19]. For $J<2$, HMVG with positive $\nu$ gives better variational
energy, hence it converges faster to the ground state. For $J>2$,
variational energy of HMVG with negative $\nu$ is also about one percent
from the ``true'' ground state energy. The agreement is not as good
for higher particle densities.

Not only the energy converged when power reaches 16, both
spin and charge structure factors have also converged as shown along several
directions in the Brillouin zone in Fig. 1(b) and 1(c).
Starting with very different variational results of GWF shown by
the dotted curves and that of HMVG shown by the solid curves, the result of
applying 16 powers of $H$ are shown by
the empty and
filled circles
using the same parameters as in Fig. 1(a).
The peak at $2{\bf k}_F=(0.6\pi,0)$ in S($\bf k$) is more pronounced
than the results of HMVG-$\nu=0.03$. Interestingly enough, a plateau-like
structure developed in D($\bf k$) for $k\ge 4k_F$. Here $4{\bf k}_F$ is
calculated by
treating all the holes, 82 of them in this case, as spinless fermions.
Both these features are observed in one dimension.

As discussed above, as far as we can tell, GWF is practically the ground
state for $J=2$ with particle densities
$18\over{100}$ and $10\over{64}$. For $J<2$ the ground state behaves
like a repulsive TLL with a very pronounced $2{\bf k}_F$ SDW correlation.
S($\bf k$) has peaks at $2{\bf k}_F$ and D($\bf k$) has its maximum at zone
boundary. This is again demonstrated in Fig. 2 by the triangles
which are the converged results for $J=0.1$. Here the particle density equals
$10\over{64}$. In this case there are two distinct $2{\bf k}_F$
vectors at
${\pi\over 4}(1,2)$ and
${\pi\over 4}(0,3)$ [22]. The former has been chosen to be examined
in detail in
Fig. 2.
Circles are results of GWF which are also the converged results for $J=2$. For
$J=3$, the converged results are denoted by squares in Fig. 2.
Just as in one dimension, $J>2$ is the region of attractive TLL. Here
SDW correlation is suppressed,
S(${\bf k}=2{\bf k}_F$) is decreased, and D($\bf k$)
has a peak at $2{\bf k}_F$.

To illuminate this cross-over behavior from repulsive to attractive TLL
as $J$ is increased above 2, we show the singlet s-wave pairing
structure factor P($\bf k$) in Fig. 3. The result of GWF shown by the
solid line is for
$J=2$. Applying eighth power of $H$ to GWF for $J=0.1$ and $J=3$,
we obtained the results shown by the empty and filled circles
respectively. P(${\bf k}=0$) is strongly suppressed
for $J=0.1$, but enhanced for
 $J=3$. Empty squares are the result of $J=4$ by applying twelfth
power of $H$ to a projected s-wave BCS trial wave function.
  The shape of the curve certainly is consistent with the
that of one-dimensional TLL. According to equation
(3)
$P(k)\sim \left|k\right|^{1\over{K_{\varrho}}}$ for small $k$.
 Comparing results of $J=3$ (filled squares) and $J=4$ (empty
squares) in figure 3, we can easily see that the cusp at $k=0$
increases with the value of $J$, hence
$K_{\varrho}$ also increases with the value of $J$. The shape of $J=4$
is unlike BCS-type state which has a delta function like P($k$).
 Another unique signature of TLL is
that P(k) has a dip at $2k_F$. Detailed examination
of this behavior is shown  in the inset in Fig. 3.
 Empty circles for $J=0.1$ clearly do not show an
increase near ${\bf k}=(\pi,\pi)$ as contrast to $J=3$ ( filled circles)
 and $J=4$ (empty squares).

In summary, we have presented energies and structure factors of the
ground states of two-dimensional $t-J$ model calculated by the power
method for 18 particles in a $10\times10$ lattice and 10 particles in
an $8\times8$ lattice. As far as we know, this is the first time that
such accurate numerical results have been obtained for such a larger
system with strongly correlated fermions.
The three structure factors, spin, charge and
pairing, show very similar behavior as in one dimension.
Within numerical accuracy GWF seems to be the ground state
for $J=2$. For $J<2$
 the  ground state is very consistent with the
picture of a repulsive TLL while an attractive TLL is
for $J>2$.

The numerical results provided above  cannot completely rule out other possible
ground states besides TLL.
But the surprising fact is that TLL provides the most natural and
consistent framework to explain ${\it all}$ the
numerical results which include many special features. A simple
random-phase-approximation [23] can explain a number of results but it
fails to explain the  pairing structure factor. This will be discussed
in detail in a future publication. We have also examined the momentum
distribution function, it is again consistent with TLL. But because
there are only several $k$ vectors in a particular direction
  it is quite difficult to make
 definite statements about the exponent. It will also be discussed in
the future.

The similarity between our results in one and two
dimensions may due to a simple reason.
For low
particle densities as we have studied here the Fermi surface is
almost a perfect circle. Hence the Fermi velocity is in the radial
direction.  We suspect this may be the reason to cause the scattering along the
radial direction to dominate and the system behaves like in one
dimension or more precisely in half a dimension.
A similar idea has been considered by Anderson [3].
The HMVG wave
function which has energy within one percent of the exact ground state
is certainly consistent with an isotropic TLL.
When the particle density is increased,  existence of other
anisotropic TLL becomes an intriguing possibility. There are evidences
that d-wave pairing becomes more important than s-wave [24].

\medskip
TKL would like to thank Materials Science Center and Department of
Physics of National Tsing Hua University for their hospitality
during his visit where part of this work is carried out.
This work was
partially supported  by
the National Science Council of Republic of China, Grant Nos.
NSC82-0511-M007-140.
\medskip

\vfil\eject

\centerline{REFERENCES}
\medskip
\item{1}{\it The Los Alamos Symposium-1989; High Temperature
Superconductivity}, edited by K.S. Bedell, D. Coffey, D.E. Meltzer,
D. Pines, and J.R. Schrieffer (Addison-Wesley, Redwood City, CA,1990).

\item{2}P.W. Anderson, Phys. Rev. Lett. \underbar{64}, 1839 (1990); M.
Ogata and P.W. Anderson, Phys. Rev. Lett. \underbar{70}, 3087 (1993),
and references therein.

\item{3}P.W. Anderson, Phys. Rev. Lett. \underbar{65}, 2306 (1990).

\item{4}S. Tomonaga, Prog. Theo. Phys., \underbar{5}, 544 (1950);
J.M. Luttinger, J. Math. Phys., \underbar{4}, 1154, (1963).

\item{5}F.D.M. Haldane, Phys. Rev. Lett. \underbar{45}, 1358 (1981);
N. Kawakami and S.K. Yang, Phys. Rev. Lett. \underbar{65}, 2309
(1990); H. Frahm, V.E. Korepin, Phys. Rev. B\underbar{42}, 10553
(1990).

\item{6}M. Ogata, M. Luchini, S. Sorella and F.F. Assaad, Phys. Rev. Lett.
\underbar{66}, 2388 (1991);
H. Shiba and M. Ogata, Prog.
Theo. Phys. Supple.\underbar{108}, 265 (1992).

\item{7}Y.C. Chen and T.K. Lee, Phys. Rev. B\underbar{47}, 11548
(1993);to appear in {\it Proceedings of Second Beijing International
Conference on High Temperature Superconductors}.

\item{8}E. Dagotto, J. Riera and A.P. Young, Phys. Rev. B\underbar{42}, 2347
(1990); K.H. Luk and D.L. Cox, Phys. Rev. B\underbar{41}, 4456 (1990).

\item{9}K.J. von Szczepanski, P. Horsch, W. Stephan, and M. Ziegler, Phys. Rev.
B\underbar{41}, 2017 (1990).

\item{10}C. Gros, R. Joynt, and T.M. Rice, Phys. Rev. B\underbar{36}, 381
(1987).

\item{11}T.K. Lee and Shiping Feng, Phys. Rev. B\underbar{38}, 11809 (1988);
T.K. Lee and L.N. Chang, Phys. Rev. B\underbar{42}, 8720 (1990).

\item{12}R. Valenti and C. Gros, Phys. Rev. Lett. \underbar{68}, 2402
(1992).

\item{13}C. Stephen Hellberg and E.J. Mele, Phys. Rev. Lett. \underbar{67},
2080 (1991).

\item{14}D.M. Ceperley and M.H. Kalos, in "Monte Carlo Methods in
Statistical Physics", edited by K. Binder (Springer-Verlag, Berlin, 1979).

\item{15}M. Boninsegni and E. Manousakis, Phys. Rev. B\underbar{47},
11897 (1993).

\item{16}J.W. Negele and H. Orland, {\it Quantum Many-Particle Systems}
(Addison-Wesley, New York, 1987).

\item{17}Shiwei Zhang and M.H. Kalos, Pyhs. Rev. Lett. \underbar{67},
3074 (1991); S. B. Fahy and D. R. Hamann, Phys. Rev. Lett.
\underbar{65}, 3437 (1990).

\item{18}C. Stephen Hellberg and E.J. Mele, Phys. Rev. B\underbar{48},
646 (1993).

\item{19}H. Yokoyama and M. Ogata, Phys. Rev. Lett. \underbar{67}, 3610 (1991).

\item{20}M. C. Gutzwiller, Phys. Rev. Lett.
\underbar{10}, 159 (1963).

\item{21}In fact appying several powers of $H$ is already enough to see the
changes.

\item{22}Other 2${\bf k}_F$ vectors are equivalent by symmetry
operations.

\item{23}J.M. Wheatley, a private discussion.

\item{24}E. Dagotto and J. Riera, Phys. Rev. Lett. \underbar{70}, 682
(1993).

\vfil\eject

\centerline{Figure Captions:}
\medskip
\item{Fig. 1} (a)Energy as a function of power, (b) spin structure
factor S($\bf k$) and (c) charge structure factor D($\bf k$)
in the $k$ space along $\Gamma$-X-M-$\Gamma$ directions
for 18 particles in a
$10\times10$ square lattice for $J=0.1$.
Empty circles are results obtained
starting from GWF and filled circles are from HMVG with $\nu=0.03$.
In (b) and (c) dashed and solid lines are VMC results for GWF and HMVG.
The
inset shows the energy as a function of power for 10 particles in an $8\times8$
lattice for $J=0.1$. Empty circles are from GWF and filled circles are
from HMVG-$\nu=0.05$. Empty squares are results for $J=2$.

\item{Fig. 2}(a)The spin structure factor S($\bf k$) along $\pi\over
4$(1,$k_y$) direction, and (b) along ${\pi}\over 4$($k_x$,2) direction
 for an $8\times8$ lattice  with 10
particles. (c) and (d) are corresponding charge structure factors D($\bf
k$). Triangles are for $J=0.1$, circles are for $J=2$ and squares for
$J=3$.  $k_x$ and $k_y$ are in unit of $\pi\over 4$.

\item{Fig. 3} The pairing correlation P($\bf k$) along (1,1) direction
for 10 particles in an $8\times8$ lattice. Solid line is the result of
GWF. Empty and filled circles are for $J=0.1$ and
for $J=3$, respectively, starting from GWF with power$=8$.
Empty squares are for $J=4$ and power$=12$ starting from a projected
s-wave BCS state. The detail for $k>2k_F$ is shown in the inset.
Along (1,1) direction, ($\pi\over 2$,$\pi\over 2$) is closest to
2${\bf k}_F$.

\end